\documentclass[aps,prl,amsmath,amssymb,reprint,floatfix]{revtex4-1}
\usepackage[utf8]{inputenc}
\usepackage{mfirstuc}
\usepackage{graphicx}
\usepackage{subfigure}
\usepackage{float}
\usepackage{gensymb}
\usepackage{amsmath}
\usepackage{color}
\usepackage{multirow}
\usepackage{natbib}
\usepackage{comment}
\usepackage[colorlinks=true, linkcolor=blue, citecolor=blue, urlcolor=blue]{hyperref}
\usepackage{breakurl}
\usepackage{url}
\newcommand{\SK}[1]{\textcolor{black}{{#1}}}

\begin{document}
%\begin{titlepage}

\title{Study of flow of crystals and deformable particles in a channel and the effective segregation of soft and hard particles}
\author{Padmanabha Bose}
\author{Smarajit Karmakar}
\email{smarajit@tifrh.res.in}
\affiliation{Tata Institute of Fundamental Research, 36/P, Gopanpally Village, Serilingampally Mandal, Ranga Reddy District, Hyderabad 500046, Telangana, India}

\begin{abstract}
\SK{Soft matters whose constituents are deformable are ubiquitous in nature especially in biological systems—including cells and their organelles—as well as in foams and emulsions. The capacity for deformation in these soft materials gives rise to a range of intriguing phenomena, such as glassy behavior without any size dispersity, cluster crystal formation, and re-entrant melting. Deformability also plays a crucial role in facilitating essential biological processes, such as the flow of blood through veins and arteries. In this work, we investigate assemblies of two-dimensional (2D) polymeric, non-overlapping rings, which mimic deformable particulates in 2D using extensive molecular dynamics simulations. The rings are confined in a rectangular channel with hard walls perpendicular to the flow direction, mimicking natural flow conditions. We analyze the flow properties of these deformable particle assemblies at two different stiffness values. To further asses the impact of deformability, we examine the same monodisperse system at higher densities for the stiffer rings, where deformation is necessary and a fluid layer emerges at the channel edges. Finally, we explore a mixture of rings with two distinct stiffnesses and observe effective segregation of soft and hard particles at small channel widths.}
\end{abstract}
\maketitle
\section{Introduction}
\SK{Soft glassy materials comprise a diverse class of systems, including colloidal suspensions\cite{Mattsson2009}, foams, emulsions\cite{C3SM51543E}, granular materials\cite{z161-sd1y}, and biological tissues\cite{PhysRevX.6.021011}. Unlike Newtonian fluids, which maintain a constant viscosity regardless of the applied shear rate ($\dot{\gamma}$), soft glassy materials display pronounced shear-dependent viscosity\cite{PhysRevE.58.738,Mizuno2024}. This non-Newtonian behavior arises from their complex microstructures and slow relaxation dynamics.
Dense colloidal suspensions exemplify these behaviors. At low shear rates, they typically exhibit shear thinning, where viscosity decreases as shear rate increases, primarily due to particle rearrangements under flow\cite{doi:10.1073/pnas.1301055110}. At higher shear rates, these systems can transition to shear thickening, where viscosity rises with increasing shear rate as particle contacts and hydrodynamic interactions play a more significant role\cite{10.1063/1.2007667}.
These phenomena are observed in both crystalline and polycrystalline solids under shear. These materials often develop distinctive features in their displacement fields, including vortices and saddle points\cite{C3SM50401H}. In crystalline systems, the origins of these singular structures are relatively well understood—they result from mechanisms such as grain rotation and grain boundary migration\cite{doi:10.1073/pnas.1210456109,PhysRevE.95.012610}. However, when similar features appear in amorphous solids, their structural origins remain elusive, presenting an open question in the field.}

\SK{Macroscopic mechanical properties of soft materials are influenced not only by particle size\cite{parakh2022ultrastablemetallicglassnanoparticles} and interparticle interactions\cite{PhysRevE.83.046106}, but also critically by particle deformability\cite{10.1063/5.0087378,Clausen_Reasor_Aidun_2011}. In complex fluids such as foams, emulsions, and cellular aggregates, the capacity of individual constituents to undergo shape changes has a profound impact on the collective mechanical response\cite{PhysRevLett.78.2020,C7SM01846K}. Increased deformability can fluidize a material, lowering its yield stress and enabling flow under smaller applied stresses\cite{PhysRevX.10.011016,bose2025understandingroleparticledeformability}. Conversely, constrained shape change may promote jamming and glassy dynamics, leading to slow relaxation and solid-like behavior\cite{bose2025understandingroleparticledeformability}. Shape fluctuations also contribute to nonequilibrium phenomena such as flow-induced segregation, as demonstrated experimentally\cite{doi:10.1073/pnas.2307061121}. These principles are especially relevant in biological flows, where soft, deformable particles—such as red blood cells—travel through vessels with widely varying diameters. The rheology of blood is highly sensitive to red blood cell deformability\cite{10.1063/5.0197208}, and pathological stiffening of cells, as seen in diseases like malaria\cite{Mohandas2012} or thalassemia, can significantly alter microcirculatory flow and tissue oxygenation. Therefore, elucidating the interplay between external shear, internal deformability, and microstructural organization in soft glassy materials is not only of fundamental interest but also essential for understanding biological function and dysfunction.} 

\begin{figure*}[!hptb]
  \raggedright
  \includegraphics[width=1.0\textwidth]{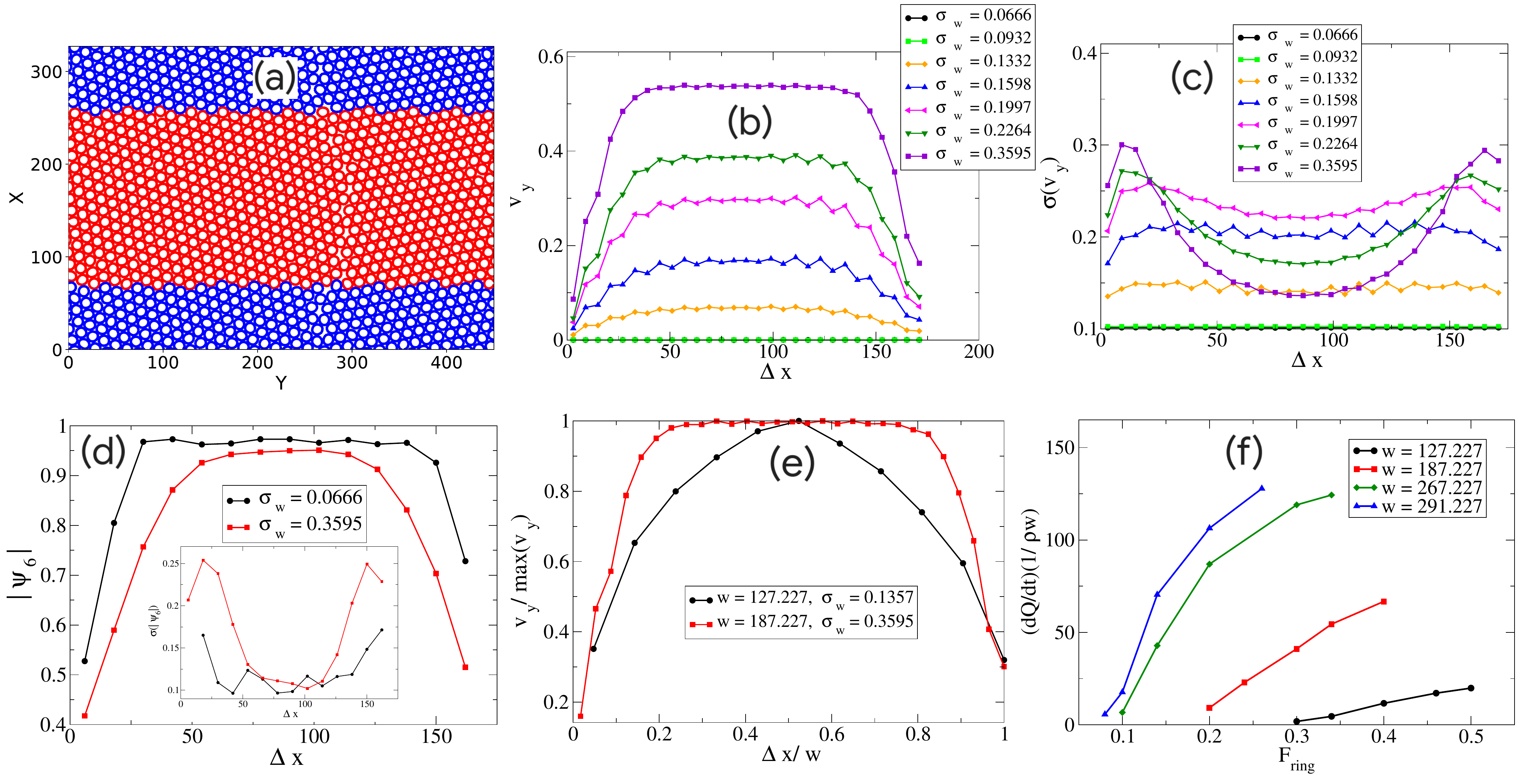}
  \caption{{\bf Flow characteristics in the hexatic phase at large stiffness, $k_\theta=150.0$.} \small(a) A typical configuration of the rings placed in a crystalline order in two dimensions. The red rings are the mobile rings, while the blue ones are the immobile ones, forming the channel walls. We show only a part of the whole channel in y direction. (b) The velocity profile of the ring CoMs along the transverse direction to flow for different values of applied-stress. For large stresses a plug-like velocity profile emerges. (c) The standard-deviation of the velocities of the ring CoMs for each bin in transverse direction to flow for same value of applied stress (d) The magnitude of hexatic-order parameter across the channel for both flowing (red) and non-flowing (black) cases. Near unity value of the hexatic order parameter at the centre of the flow signifies strong hexatic ordering even during a flow at the centre during plug-flows. Inset shows the standard deviation of the same quantity across bins and there is a strong variation in ordering near walls signifying higher strain. (e) The plug-like and parabolic flow profiles for two different widths of channel at two different applied stresses. The axes are scaled appropriately. (f) The flow of mass with forcing for different values of channel widths}
  \label{fc}
\end{figure*}

\section{Model and Methods}
\SK{In this paper, we aim to bridge the gap between hard and soft matter systems by employing two-dimensional ring-polymer assemblies. Due to their deformability, these assemblies exhibit both crystalline and amorphous behavior as ring stiffness decreases \cite{bose2025understandingroleparticledeformability}. Our primary objective is to simulate flow in a two-dimensional channel by applying a constant force to each particle \cite{PhysRevLett.60.1282}. This approach replicates the effect of a constant pressure gradient, resulting in a uniform force on each ring within a layer transverse to the flow direction. Although the applied force is constant, interactions between the channel walls and particle motion generate a non-uniform shear profile across the channel.} 

\SK{The preparation protocol is as follows. Each ring consists of $20$ monomers connected by FENE bonds, and inter-monomer repulsive WCA potential prevents overlap \cite{bose2025understandingroleparticledeformability}. We simulate multiple ring polymers in two dimensions within a rectangular box at low temperature ($T = 0.2$) and two stiffness values ($k_\theta = 20.0, 150.0$). Rings in the boundary layer, around $5$ layers( which is changed to change the width of the channel), are immobilized to form the channel walls, while a constant force is applied to the remaining mobile rings. A temperature rescale thermostat maintains constant flow temperature. The steady-state velocity profile results from the balance between the thermostat and the applied force.} 

\SK{For simulations, we use an assembly of $5041$ rings, each with a diameter of approximately $11$. In the initial study of softness effects at hexatic density, the box size is approximately $327 × 2166$. For subsequent investigations involving increased density and segregation of soft and hard particles, the box size is approximately $290 × 2166$. The constant force is applied to all mobile particles in the $\hat{y}$ direction.}

\section{Flow Characteristics of poly-crystalline assembly through a channel}
\begin{figure*}[!hptb]
  \raggedright
  \includegraphics[width=1.0\textwidth]{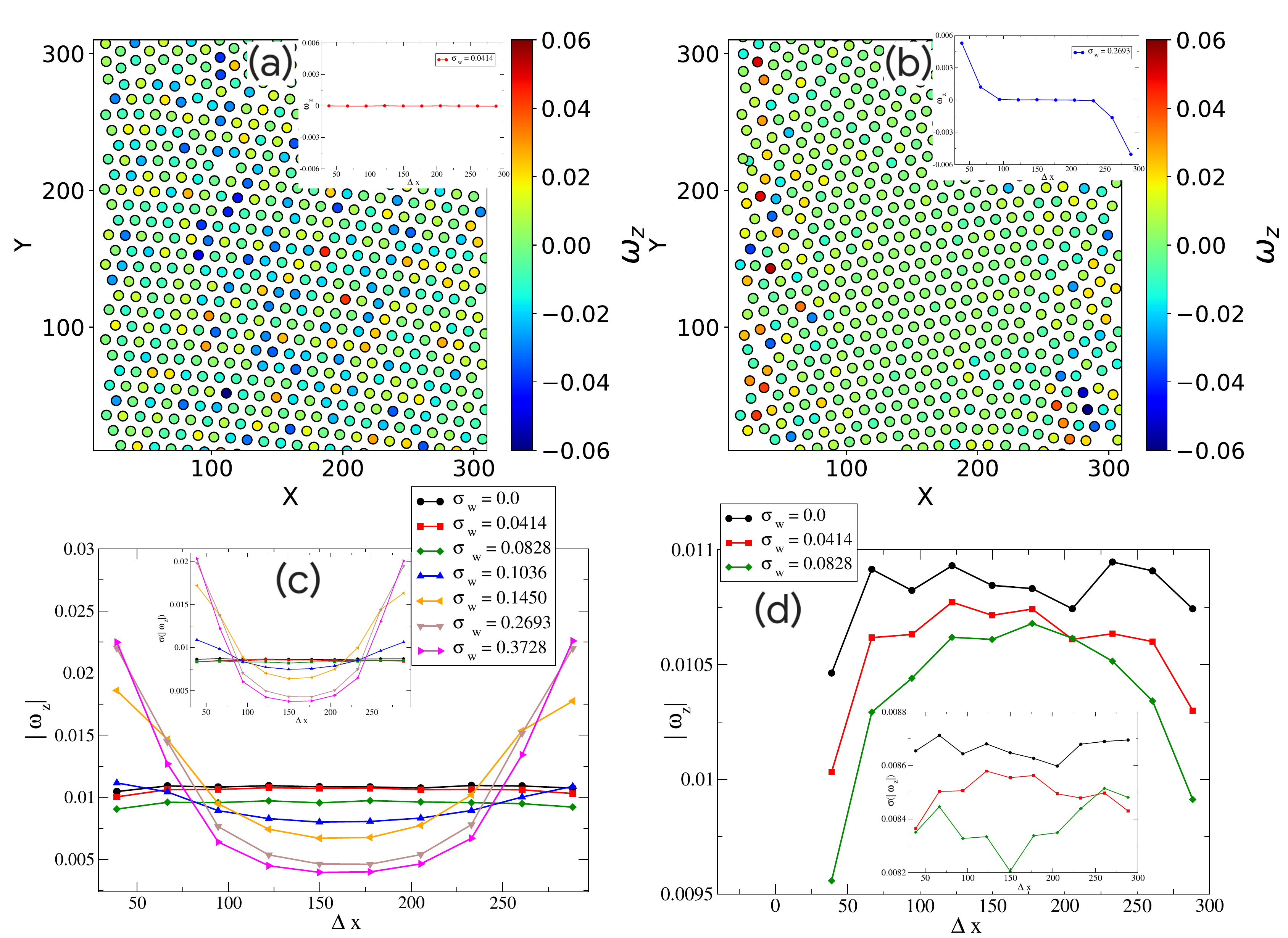}
  \caption{{\bf The rotational characteristics in the hexatic phase.} \small(a) A snippet of the channel when there is no flow through the channel with the Ring CoMs represented by points. The inset shows the averaged variation of $\omega_z$ in the transverse direction. (b) The same snippet when there is a flow through the channel. The inset clearly represents the change of rotational direction across the channel due to the flow. (c) The absolute value of $\omega_z$ variation across channel with ($\sigma_w\le0.0828$) and without flow ($\sigma_w\ge0.1036$). The inset shows the standard deviation of $\omega_z$ across the channel (d) $|\omega_z|$ across the channel for three different values of forcing are shown. For $\sigma_w\le0.0828$, the system does not flow, and one can clearly see the indication of jamming as $|\omega_z|$ decreases with increase in $\sigma_w$ which indicates less rotation of the rings with increase in forcing which indicates a more jammed state. The inset shows the standard deviation of $|\omega_z|$ across the channel width.}
  \label{omega}
\end{figure*}
\begin{figure*}[!hptb]
  \raggedright
  \includegraphics[width=1.0\textwidth]{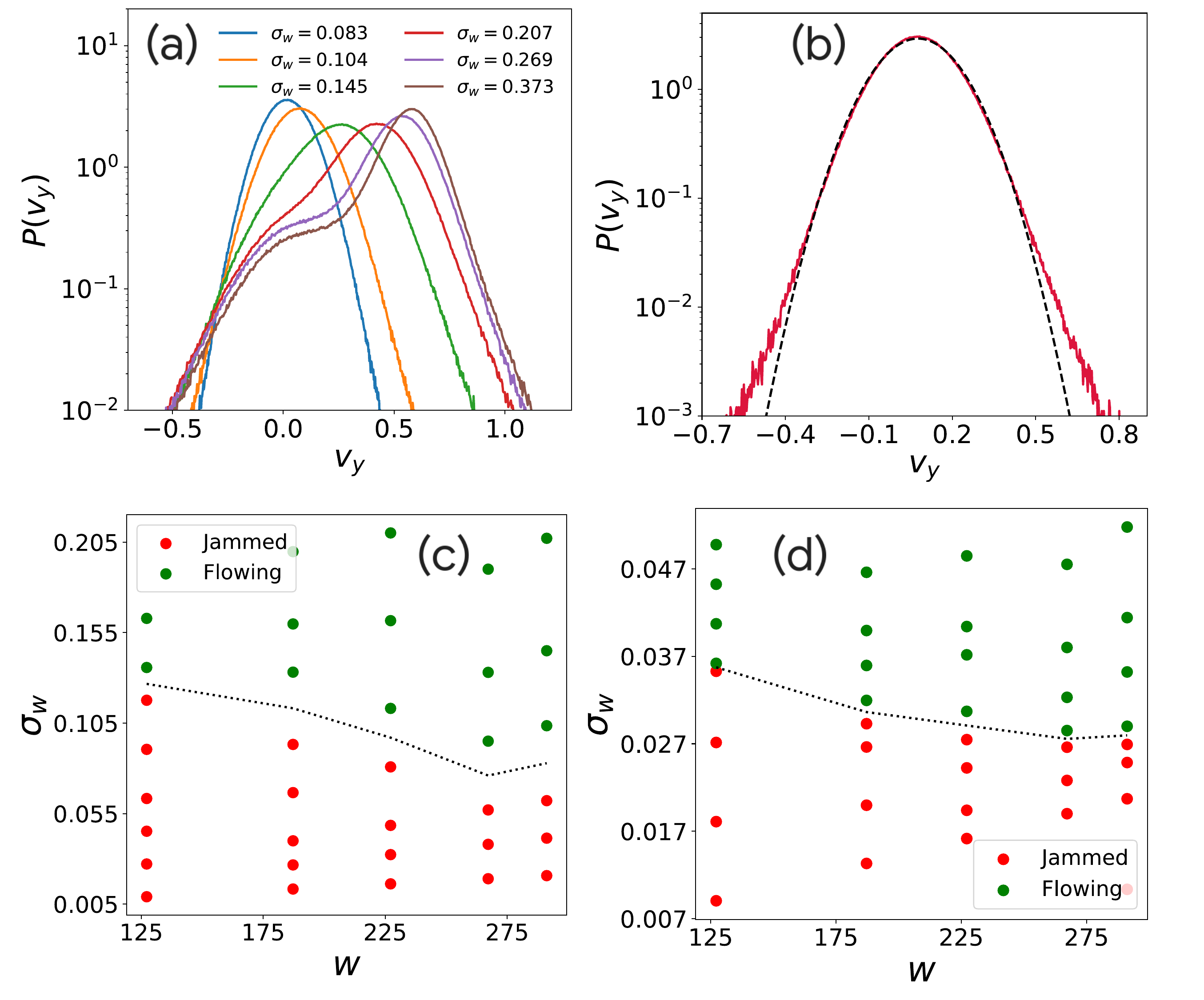}
  \caption{{\bf The velocity profiles and jamming-unjamming transitions on application of stress} \small(a) Typical velocity distribution along the flow direction with humps appearing at a +ve $v_y$ in flowing states, (b) Velocity distribution for $\sigma_w=0.104$. The mean of the distribution is greater than 0, and the distribution is skewed towards $+v_y$, (c) The phase-portrait of $\sigma_w$ vs w for $k_\theta=150.0$, the red dots represent jammed states and the green dots represent flowing states (d) The same portrait for ring-stiffness, $k_\theta=20.0$, at the same density. Note the reduction of stress-values in this case for the same channel widths.}
  \label{fj}
\end{figure*}
\SK{Increasing the stiffness value ($k_{\theta}=150.0$) at a density where the rings are closely packed yet undeformed results in the formation of a crystalline structure by the ring centers of mass (CoMs). The system is characterized by applying forces to each monomer along the channel direction. Fig.\ref{fc}(a) presents a snapshot of the entire channel, where blue rings are immobile and constitute the channel wall, while red rings are mobile. Fig.\ref{fc}(b) displays a representative velocity profile for the flow under varying forcing values, defined as $\sigma_w=\frac{w.\rho_{COM}.F_{COM}}{2}$, where $w$ is the channel width, $\rho_{COM}$ is the density of rings, and $F_{COM}$ is the total applied force on each ring. When the forcing remains below a certain threshold, the rings exhibit only minor fluctuations and do not move significantly \cite{PhysRevE.77.011504}. Once the forcing exceeds this threshold, the rings begin to move, forming a parabolic flow profile at low $\sigma_w$ values. At higher forcing, the profile transitions to a plug-like shape with a constant velocity at the center. This transition from a smooth-
parabolic to a plug-like profile with increasing forcing is reflected in the standard deviation of velocities within these bins. In plug-like flow, the central mass of rings moves collectively, resulting in minimal velocity deviations among the CoMs in these bins, which reduces the standard deviation as shown in Fig.\ref{fc}(c). In contrast, during smooth-parabolic flow, heterogeneity in the flow maintains a constant standard deviation across the channel width. Beginning from a near-crystalline arrangement, if the central rings move collectively, these rings are expected to retain hexatic order as computed from
\begin{equation}
\psi_6=\frac{1}{N_n}\sum_{j=1}^{N_n}e^{i6\theta_j}, 
\end{equation}
Here, $N_n$ denotes the number of neighbors for each ring, determined using their respective CoMs. This is evident from the red curve in the main panel of Fig.\ref{fc}(d). In the non-flowing case, the hexatic phase exhibits a peak that is more pronounced than in the flowing case. The variation of the hexatic phase in each bin indicates that, during flow, the hexatic phase changes near the arms of the plug due to significant rearrangements in that region. In Fig.\ref{fc}(e), both the width and velocity axes are scaled to their maximum values. For channels with smaller widths and lower stress, the velocity profile is parabolic, whereas for larger widths and stronger forcing, the profile reverts to a plug-like shape. Fig.\ref{fc}(f) presents the mass flow at different channel widths. For the same value of forcing per ring, larger channels yield at very low forcing values, while smaller channels require a significantly higher force to initiate flow.}

\begin{figure*}[!hptb]
  \raggedright
  \includegraphics[width=1.0\textwidth]{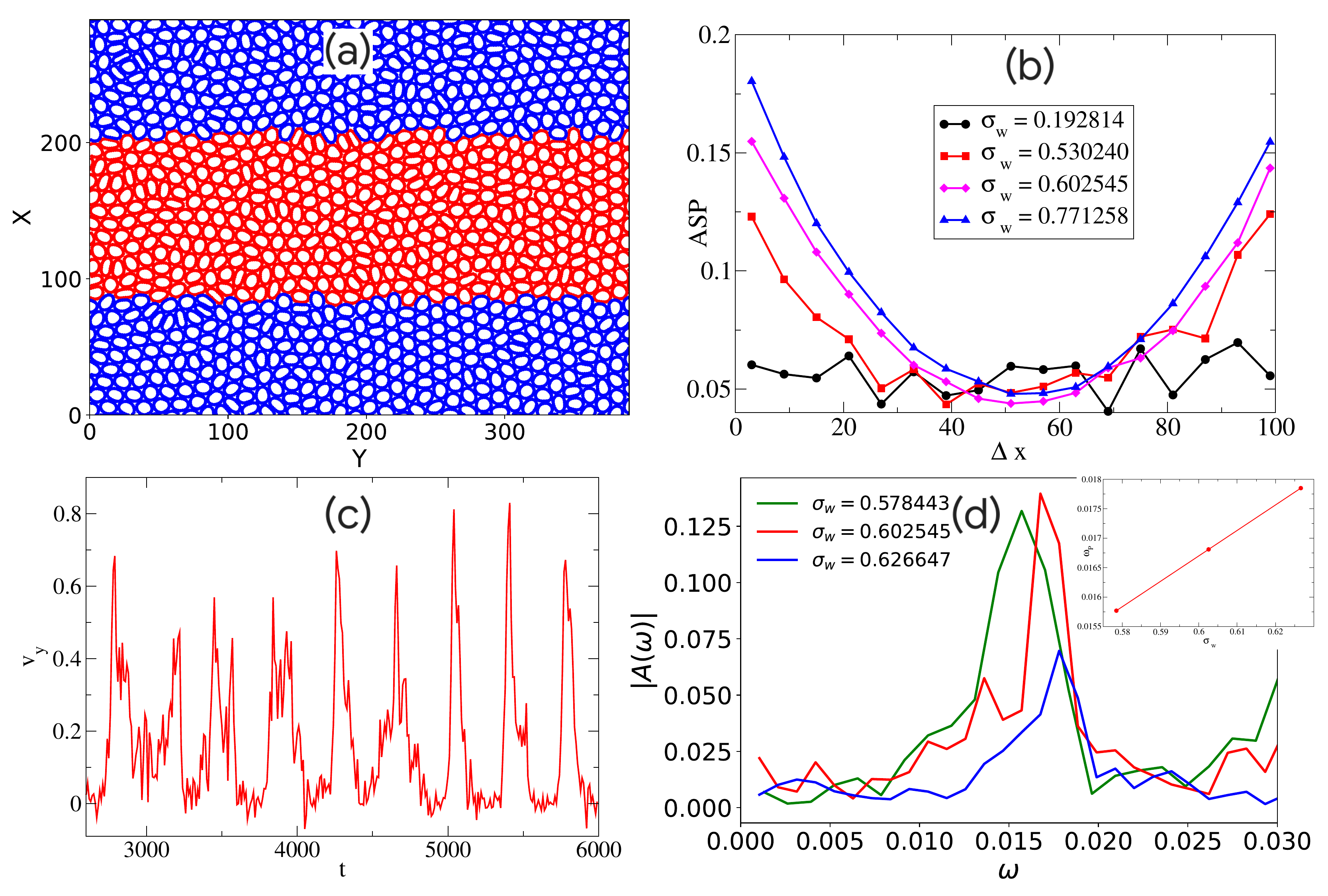}
  \caption{{\bf Travelling-waves in a dense system of mono-disperse ring polymers} \small(a) Snapshot of a section containing mono-disperse stiff ($k_\theta=150.0$) rings in a channel. The red rings in the central region represent mobile rings, and the blue rings form the channel walls. (b) Variation in the asphericity of the rings across the channel. The system flows for $\sigma_w\ge0.6025$. For stresses larger than the flowing stress, the asphericity parameter starts to develop a U-shape characteristic with channel width, $\Delta x$ (see text). (c) The temporal velocity variation in a fixed-square patch. Note the near-periodic spikes in the y-velocity of the rings. (d) The fast Fourier Transform (FFT) of the time signals for three different values of $\sigma_w$ at which the periodic spikes (travelling waves) appear. The inset shows the near-linear increase in the frequency of the waves (taken from the dominant peak position in the FFT of the signals) with an increase in $\sigma_w$.}
  \label{tw}
\end{figure*}
\SK{A previous study \cite{PhysRevLett.124.158003} on the flow of crystalline systems in a channel predicted behavior similar to that observed in the present work. However, the particulate nature of the systems in that study prevented probing the angular movement of individual rings, which is expected in flows due to the differential movement of each layer. This limitation is not present in the current system. The angular dynamics are investigated, and key observations are summarised in Fig.\ref{omega}. Fig.\ref{omega}(a) presents a snapshot of the channel before the onset of flow ($\sigma_w=0.0414$). Due to thermal motion, the rings rotate about their centers without significant displacement. The angular velocity of each ring is calculated using the positions of the 1st and 10th monomers, with data binned at intervals of $6.0$. Averaging over multiple snapshots yields an average angular velocity.
\begin{equation}
\omega_z=\frac{1}{N_{snap}.N_{CoM}}\sum_{i=1}^{N_{snap}}\sum_{j=1}^{N_{CoM}}\frac{\Delta\theta_{i,j}}{\Delta t_i}
\end{equation}
The average angular velocity is found to be approximately zero. Fig.\ref{omega}(b) displays the system in a flowing state, with forcing applied along the positive y-direction. Averaging over multiple configurations reveals that rings on the left side of the channel rotate anti-clockwise, while those on the right rotate clockwise, consistent with expectations. To minimize thermal contributions to ring rotation, the absolute value of the angular velocity for each ring is plotted in Fig.\ref{omega}(c). Prior to yielding, thermal motion results in a nearly constant $|\omega_z|$ across bins, with minor decreases at the boundaries due to immobile rings. Upon initiation of flow, a dip in $|\omega_z|$ appears at the center of the channel, which becomes more pronounced with increased forcing. Simultaneously, $|\omega_z|$ increases at the boundaries as forcing intensifies. The central dip in $|\omega_z|$ indicates a more stressed central plug region that moves cohesively, with stress increasing as forcing rises. The variation of $|\omega_z|$ across bins supports this interpretation. Fig.\ref{omega}(d) presents the same quantity before yielding. Increased forcing prior to yield is expected to further stress the rings, resulting in a decrease in $|\omega_z|$, as observed in the plot. The inset illustrates the deviation of this quantity, reinforcing the same conclusion.}

\SK{When a constant force is applied to each ring, the center-of-mass (CoM) velocity distribution at high stress exhibits a pronounced peak at a positive value of $v_y$, as shown in Fig.\ref{fj}(a). At lower forcing, where the system has only just begun to flow, deviations from the Gaussian distribution of $v_{y}$ appear in the distribution tails, as illustrated in Fig.\ref{fj}b. Notably, there is a slight shift in the mean of the distribution. As previously discussed, a finite threshold stress exists beyond which the system transitions to a flowing state. The phase diagram for $k_{\theta}=150.0$ presents the distinction between flowing and jammed states. The average of the last jammed and first flowing states for each channel width is used to delineate the boundary between these regimes, indicated by a dotted line. A similar analysis at the same density and temperature but with lower stiffness ($k_\theta=20.0$) reveals a similar increase in yield stress with channel width, although the yield stress is significantly reduced, consistent with previous findings \cite{D4SM01069H}. It is important to note that between the continuously flowing and fully jammed states, a regime characterized by intermittent dynamics has been reported \cite{PhysRevLett.124.158003}. In the present state diagram, these intermittent states are classified as part of the flowing regime.}

\SK{Reducing the stiffness of the rings enables their movement under lower applied force, although the rings do not undergo permanent deformation. To have more deformation in some of the rings, the system density is increased while maintaining a constant $k_\theta=150.0$. Figure \ref{tw}a illustrates the initial state of the system, where red rings are mobile and blue rings are immobile. At very low applied stress, the distribution of the asphericity parameter (ASP) is not significant across the transverse direction. We define ASP as 
\begin{equation}
ASP=\bigg(\frac{\lambda_1^2-\lambda_2^2}{\lambda_1^2+\lambda_2^2}\bigg)^2,
\end{equation} 
where $\lambda_1$ and $\lambda_2$ are the largest and smallest eigenvalues of the shape matrix. The shape matrix is a two-dimensional matrix, defined as 
\begin{equation}
S_{pq}=\frac{1}{N}\sum_{i=1}^{N}(r_p^i-R_{p}^i)(r_q^i-R_{q}^i),
\end{equation}
for each ring where $p$ and $q$ are the cartesian coordinates in two dimensions, and they take two values $x$ and $y$. $N$ is the number of monomers ($N = 20$) in this case. $R_{p}^i$ is the centre of mass position, respectively. Immediately prior to the onset of flow, the asphericity parameter develops a U-shaped profile along the transverse direction, with rings at the center remaining largely undeformed and those at the edges becoming more elongated. As flow commences, the asphericity parameter approaches zero at the center but increases near the walls, where shear strain is greatest, resulting in elongated rings as shown in Fig.\ref{tw}(b).}

\SK{In Ref.\cite{PhysRevLett.124.158003}, it is reported that the system displays density-wave-like behavior within a narrow stress range between flowing and jammed states. Travelling waves have also been observed in gravity \cite{Ellingsen_2010}, traffic flow \cite{PhysRevE.61.3564}, and channel flow studies \cite{PhysRevLett.91.224502}. To analyse this behaviour, we select a $54.4\times54.4$ region at the flow center and measure its average center-of-mass velocity at each fixed timestep. If a travelling wave of cooperative ring movement crosses the periodic boundary, the velocity pattern should reflect this, with periodic velocity peaks appearing over time, as shown in Fig.\ref{tw}(c). To quantify this periodicity, we perform a fast Fourier transform on the velocity signal for three yield-stress values where travelling waves are present. We observe a distinct peak in the spectrum, as shown in Fig.\ref{tw}(d). As yield-stress increases, the peak position in the spectrum also increases for all three values. Plotting yield-stress against peak frequency yields a straight line, as shown in the inset of Fig.\ref{tw}(d).}

\begin{figure*}[!hptb]
  \raggedright
  \includegraphics[width=1.0\textwidth]{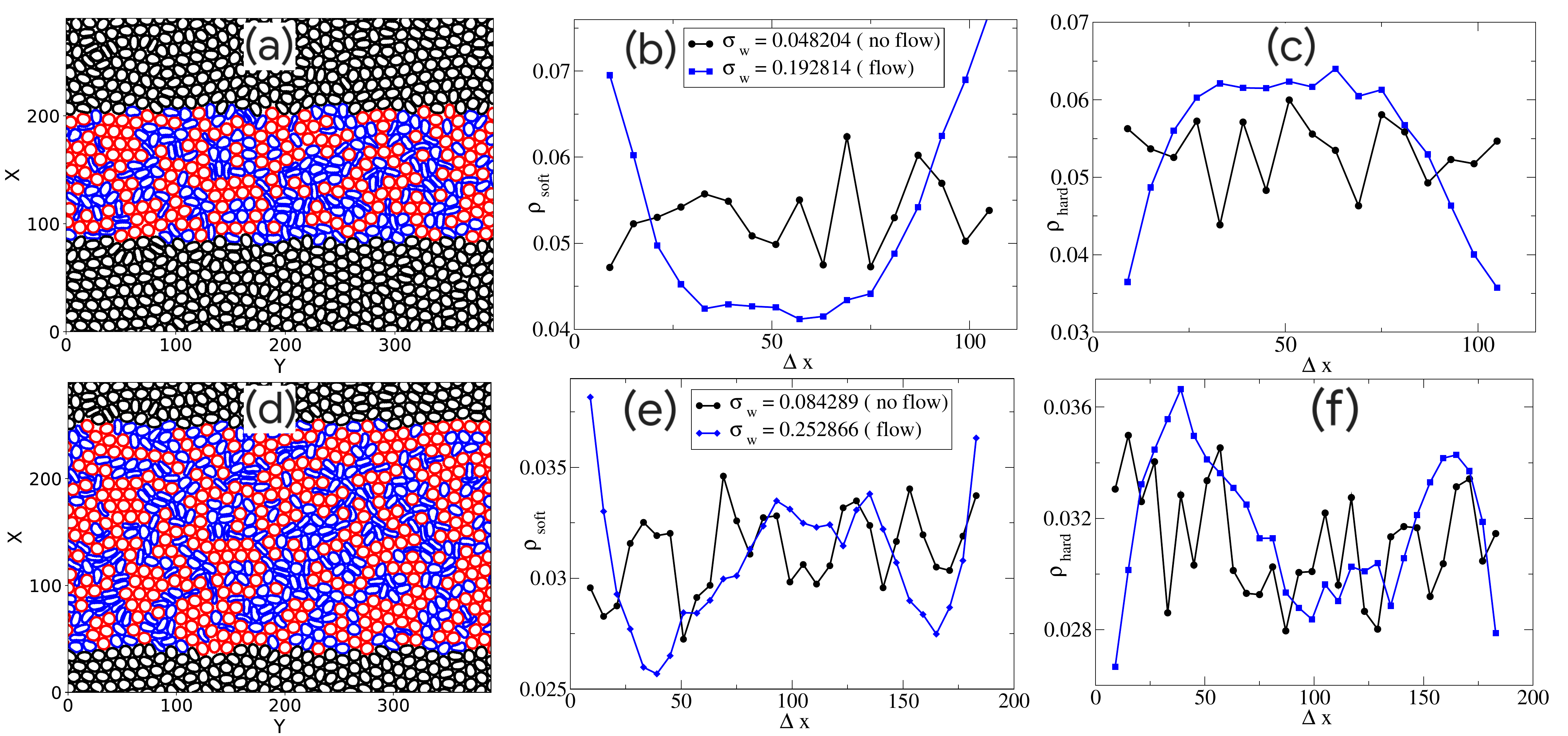}
  \caption{{\bf Separation of soft and hard rings based on channel width} \small(a) Snapshot of a section containing bi-disperse stiff ($k_\theta=20.0$ and $150.0$) rings with a channel width of $\approx130$. Hard rings are coloured red, and softer ones are coloured blue. The wall is represented as black rings. (b),(c) The soft rings get pushed towards the walls while the hard rings move towards the centre of the flow (blue curve). For comparison, without flow, the variation of the hard and soft ring density is almost constant (black curve). (d) Snapshot of the channel with wider width $\approx210$, (e),(f) Margination effect observed with softer rings clumping at the centre and harder rings clumping near the walls due to hydrodynamic lift effect (blue curves). The variation of the density of the soft and hard rings shows no such effect in the absence of flow. One point to note is that increased forcing ($\sigma_w$) doesn't change the intensity of this separation in a significant way in both these cases. }
  \label{mar}
\end{figure*}

\section{Sorting Hard and Soft Rings using Channel Flow}
\SK{The observation that shear strain is highest at the channel walls can be used to separate soft and hard particles. In a mixture of soft and stiff rings, the softer rings tend to migrate toward the channel edges, where the strain is greatest, because it costs less energy to deform a soft ring than a stiff one. To test this concept, we created a 50:50 mixture of rings with two different stiffnesses ($k_\theta=20.0$ and $150.0$), but with the same diameter. Fig.\ref{mar}(a) shows immobile rings in black, stiff rings in red, and soft rings in blue for a narrow channel. Fig.\ref{mar}(b) displays the density profiles in both flowing and jammed states. In the jammed state, there is no clear separation between soft and stiff particles. However, once flow begins, the soft rings move toward the channel walls, while the stiff rings cluster near the center, as illustrated in Fig.\ref{mar}(c). When the channel width is increased, as shown in Fig.\ref{mar}(d), the particles have more space to rearrange. In this wider channel, the elongated soft rings are drawn toward the center by the higher flow velocity (a phenomenon known as hydrodynamic lift), while the stiff rings are pulled toward the walls, reversing the previous pattern (see Figs.\ref{mar}e and \ref{mar}f). This effect is also observed in blood flow, where it is called margination: stiffer white blood cells (WBCs) and platelets migrate toward the vessel walls, while softer red blood cells (RBCs) move toward the center \cite{PhysRevLett.109.108102,C3SM52860J}. Notably, the extent of separation between soft and stiff rings does not depend on the total applied stress once flow begins. There is a critical channel width at which margination occurs, but the tendency for separation does not depend on the channel width in either regime.}

\section{Conclusion and Outlook}
\SK{This study systematically investigates the role of particle softness on the flow behaviour of confined ring polymers in channel geometry. The findings reveal a complex interplay among applied stress, confinement, structural order, and particle deformability. At small applied stresses and narrow channel widths, the transverse velocity profile is parabolic, characteristic of laminar flow. Under these conditions, the system largely preserves its hexatic order. When the channel is sufficiently wide to permit plug-like flow, spatial heterogeneity emerges: the hexatic phase remains intact near the channel center but is disrupted near the walls, where particle rearrangements are most pronounced, consistent with earlier findings \cite{PhysRevLett.124.158003}.}

\SK{It is interesting to note that internal stress can be estimated based on the absolute value of the angular velocity, $|\omega_z|$, of the rings computed between successive snapshots under applied stress. The results indicate that $|\omega_z|$ decreases with increasing stress in the pre-yield regime. After yielding, the system exhibits strong spatial differentiation: the central region shows reduced $|\omega_z|$, indicating a highly stressed configuration, while the edge regions display enhanced angular velocities due to increased local rearrangements. At higher applied stresses, the translational velocity distribution along the flow direction develops a pronounced peak corresponding to the mean flow velocity, deviating significantly from a Gaussian distribution. A threshold stress separates jammed and flowing states, and this threshold increases as the channel width decreases, in agreement with results reported in \cite{PhysRevLett.124.158003}.}

\SK{Introducing softness at fixed density by decreasing $k_\theta$ to $20.0$ results in a marked reduction in the yield stress, consistent with previous results in similar systems \cite{D4SM01069H}. At higher densities in the monodisperse system with stiffer rings ($k_\theta = 150.0$), rings near the channel walls deform into ellipsoidal shapes at the onset of flow. This deformation, driven by higher strain near the boundaries, facilitates stress relaxation and promotes flow. Near the onset stress, the system exhibits jerky-flow behavior. Tracking the center-of-mass velocities of rings in a central patch reveals periodic oscillations within a finite window of applied stress. The frequency of these oscillations scales linearly with the applied stress. Beyond this regime, the system transitions to steady, continuous flow.}

\SK{Moreover, for a $50:50$ bidisperse mixture of hard and soft rings, we find confinement-dependent segregation. In narrow channels, softer rings preferentially migrate toward the walls, where rearrangements are elastically more favorable. When the channel width exceeds a threshold, this trend reverses: hydrodynamic lift arising from the asymmetric transverse flow profile drives the softer rings toward the center, reminiscent of margination effects in blood flow\cite{}. In both regimes, the degree of separation remains largely insensitive to the applied stress, and the effect of channel width beyond the threshold is minimal.}

\SK{Incorporating activity into the rings might provide us a controlled framework to model biological transport phenomena such as blood flow \cite{ASHTON2007203} and amoeboid motion \cite{10.1371/journal.pone.0027532}, both in the presence and absence of background flow. Motivated by recent interest in the correspondence between active forcing and oscillatory shear \cite{Sharma2025}, a promising direction for future research is to explore whether periodic forcing applied to a glassy phase can induce annealing effects analogous to those observed under oscillatory deformation.} 

\section{Acknowledgements :}
We acknowledge funding from intramural funds at TIFR Hyderabad, provided by the Department of Atomic Energy (DAE) under Project Identification No. RTI 4007. SK acknowledges Swarna Jayanti Fellowship grants DST/SJF/PSA01/2018-19 and SB/SFJ/2019- 20/05 from the Science and Engineering Research Board (SERB) and Department of Science and Technology (DST). Most computations are performed using the HPC clusters procured through Swarna Jayanti Fellowship grants DST/SJF/PSA01/2018-19 and SB/SFJ/2019-20/05. SK also acknowledges the research support from the MATRICES Grant MTR/2023/000079, funded by SERB.
\bibliographystyle{apsrev4-2}
\bibliography{main}

\end{document}

%\preprint{APS/123-QED}
%\begin{titlepage}
\title{Supplementary material :}% Force line breaks with \\
%\thanks{A footnote to the article title}%

\author{Padmanabha Bose}

\author{Smarajit Karmakar}
\affiliation{Tata Institute of Fundamental Research Hyderabad}

\maketitle

\section{Mean-squared-displacement of monodisperse assembly of rings in a channel with and without flow :}

\begin{figure}[!htb]
  \raggedright
  \includegraphics[width=0.85\columnwidth]{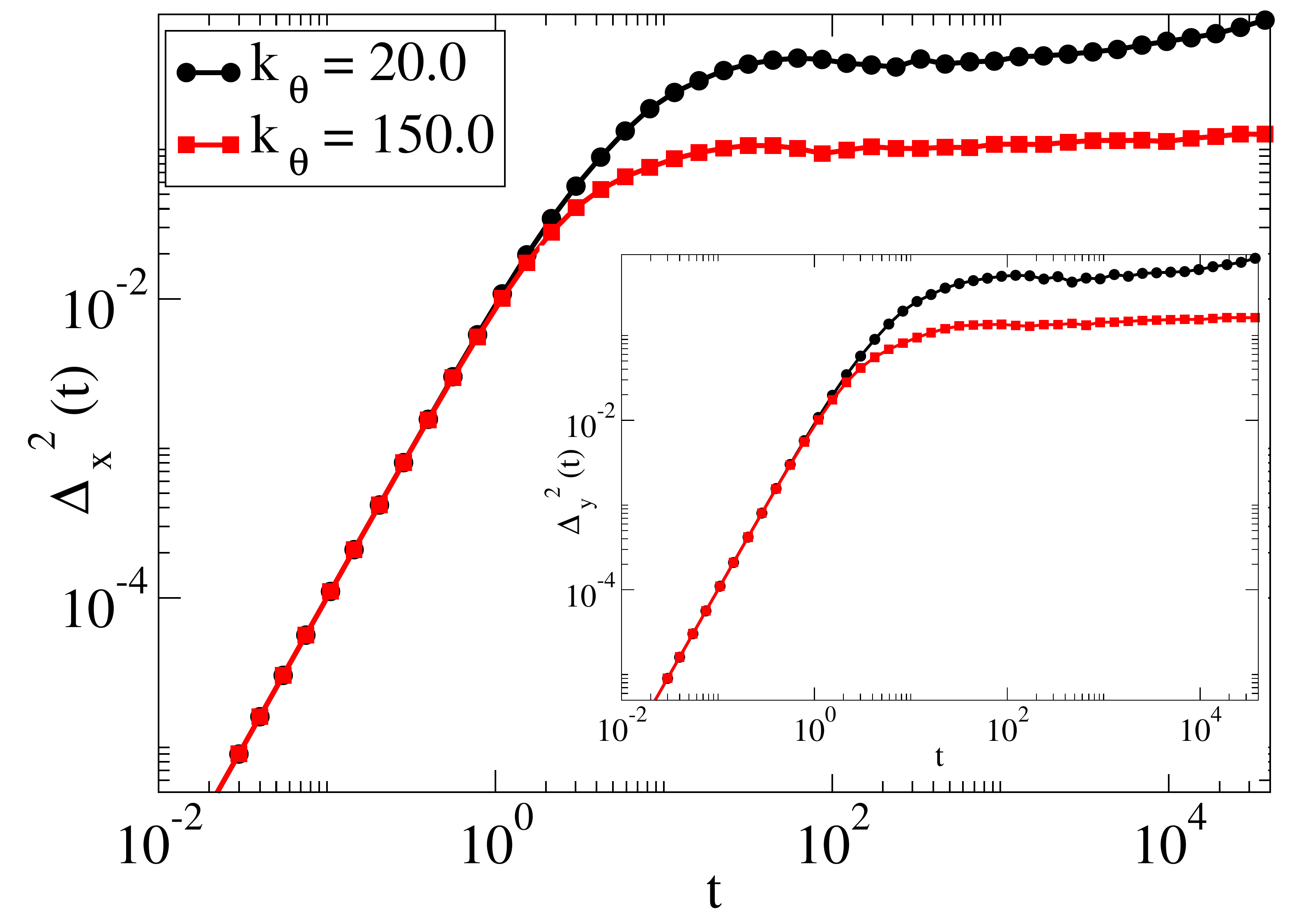}
  \caption{Mean squared displacement in both the flow and transverse directions without forcing for different values of $k_\theta$s( $k_\theta=20.0$ and $k_\theta=150.0$).}
  \label{nf}
\end{figure}

\begin{figure}[ht]
\centering
\includegraphics[width=\textwidth]{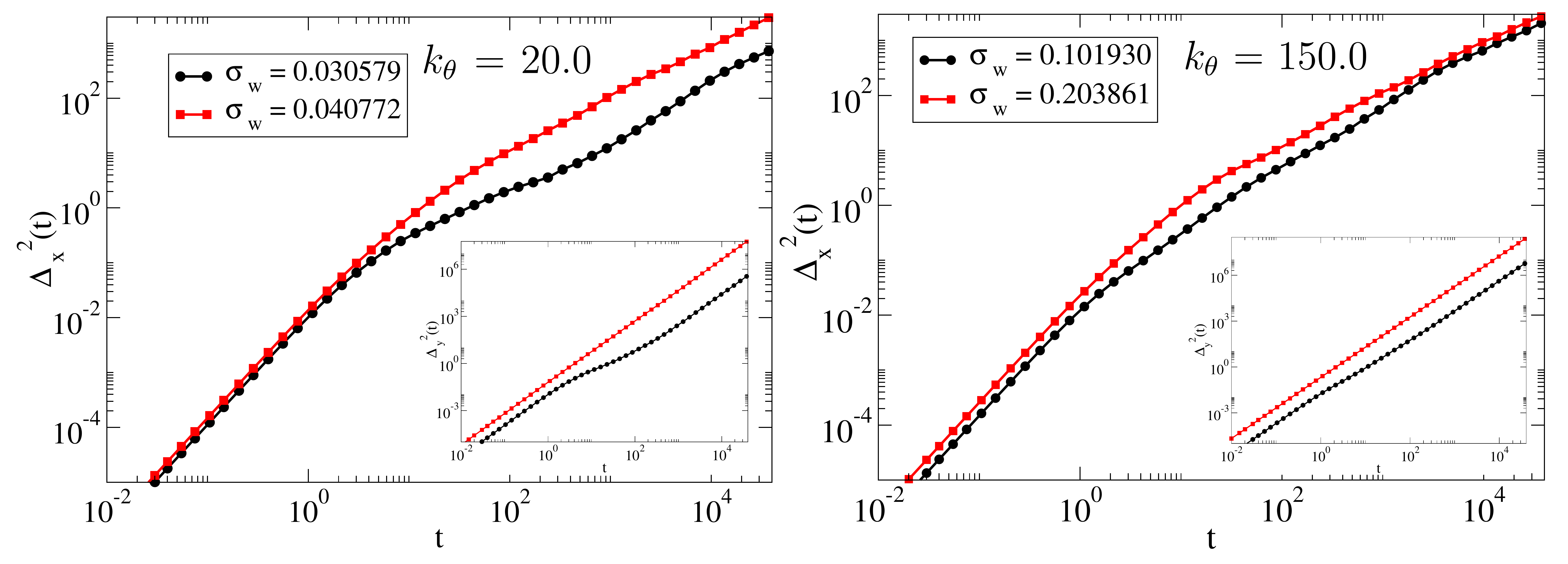}
\caption{The left plot is the MSD in the x-direction for $k_\theta=20.0$ while the inset shows the MSD for the same stress values in y direction. The right plot is the MSD in the x-direction for $k_\theta=150.0$ while the inset is the MSD in the y direction i.e along the flow direction. Note the increase in the stress values for higher $k_\theta$ values. Also note the diffusive behaviour in the long-time x-direction MSD vs the y-direction MSD which is ballistic}
\label{wf}
\end{figure}

When there is no forcing in the system, at the low temperature( T=0.2) and density, the rings behave as jammed rings and the mean squared displacement( MSD=$\langle(\Delta r)^2\rangle$) shows a saturation of MSD both along and transverse to the channel( the MSD along the x direction is slightly lower due to presence of hard walls in that direction)( see \ref{nf})( this is at the same density as that of the initial part of analysis with the hexatic centre). As soon as flow starts, the rings in the flow direction move along the channel with a steady state velocity, which causes the MSD along the channel to increase ballistically. Though there is no forcing in the transverse direction, the flow also causes the facilitation of the rings in the transverse direction, which makes the MSD grow diffusively in the transverse direction till the hard wall is encountered by the particle( see \ref{wf}). The MSD is more for softer particles in both cases due to obvious reasons.

\onecolumngrid

\twocolumngrid